\documentclass{aastex701}

\newcommand{\beq}{\begin{equation}}
\newcommand{\eeq}{\end{equation}}

\newcommand{\Ms}{M_*}
\newcommand{\Msun}{M_\odot}
\newcommand{\kmps}{km~s$^{-1}$}
\newcommand{\hi}{H{\sc i}}
\newcommand{\MHI}{\langle{M_{\scriptsize{\textrm{H\textsc{i}}}}}\rangle}

\newcommand{\hii}{H{\sc i}\,21cm}

\begin{document}

\title{The GMRT CAT$z$1-COSMOS Survey: H{\sc i} 21\,cm emission from star-forming galaxies at $z\approx1$ in the COSMOS field}

\author{Balpreet Kaur} 
\affiliation{Inter-University Centre for Astronomy and Astrophysics, Pune University, Pune 411007, India}
\affiliation{National Centre for Radio Astrophysics, Tata Institute of Fundamental Research, Pune University, Pune 411007, India}
\email{balpreet.kaur@iucaa.in}

\author{Nissim Kanekar} 
\affiliation{National Centre for Radio Astrophysics, Tata Institute of Fundamental Research, Pune University, Pune 411007, India}
\email{nkanekar@ncra.tifr.res.in}

\author{Aditya Chowdhury}
\affiliation{National Centre for Radio Astrophysics, Tata Institute of Fundamental Research, Pune University, Pune 411007, India}
\email{aditya.chowdhury@outlook.com}

\author{Jayaram N. Chengalur} 
\affiliation{National Centre for Radio Astrophysics, Tata Institute of Fundamental Research, Pune University, Pune 411007, India}
\email{chengalur@ncra.tifr.res.in}

\begin{abstract}

We report a 216-hr upgraded Giant Metrewave Radio Telescope (GMRT) Band-4 $550-850$~MHz observation of the COSMOS field, using 7 GMRT pointings to cover an area of $2.4$~sq.~deg., with on-source times of $67$~hrs in the central pointing, and $89$~hrs distributed across the other six pointings. We stacked, at a spatial resolution of 90~kpc, the H{\sc{i}}~21\,cm emission signals from 8,122 star-forming galaxies with accurate spectroscopic redshifts and lying within the FWHM of the GMRT primary beam at their redshifted H{\sc{i}}~21\,cm frequencies. We detect the stacked H{\sc{i}}~21\,cm signal from the 8,122 galaxies at $4.2\sigma$ significance, obtaining an average H{\sc{i}} mass of $\langle{M_{\scriptsize{\textrm{H\textsc{i}}}}}\rangle=(15.5\pm3.7)\times10^9\,\textrm{M}_\odot$. This is only the second galaxy population at $z\approx1$ with a measurement of the average H{\sc{i}} content; the other being the blue star-forming galaxies in the DEEP2 fields. We also detected the median 1.4~GHz radio continuum emission signal from the 8,122 galaxies by stacking the GMRT 680~MHz continuum images, obtaining a weighted-median star-formation rate (SFR) of ${8.90\pm0.92}\,M_\odot$~yr$^{-1}$. We measure a characteristic H{\sc{i}}-to-stellar mass ratio of $\langle{M_{\scriptsize{{\textrm{H{\sc i}}}}}}\rangle/\langle{M_\star}\rangle=(1.15\pm0.27)$ and a characteristic H{\sc{i}} depletion timescale of $\langle{M_{\scriptsize{{\textrm{H{\sc i}}}}}}\rangle/\textrm{SFR}=(1.74\pm0.45)$~Gyr for star-forming galaxies at $\langle{z}\rangle=1.081$ in COSMOS, both consistent with earlier measurements for the blue, star-forming DEEP2 galaxies at $z\approx1$. Our results support the evidence from the DEEP2 survey that the observed decline in the SFR density at $z<1$ is due to insufficient accretion of atomic gas from the intergalactic medium to replenish the gas reservoir consumed in the process of star-formation.     

\end{abstract}


\section{Introduction}
\label{sec:intro}

The cosmic evolution of galaxies is an interplay between complex competing physical processes constituting the baryon cycle. The key processes are the infall of gas onto galaxies to form neutral atomic hydrogen (H{\sc i}), the cooling of \hi\ to yield the cold atomic phase, the conversion of neutral hydrogen from the cold atomic state to the molecular state (H$_2$), the fragmentation and collapse of molecular clouds to form stars, and feedback from the supernovae and stellar winds that drive metals into the interstellar medium (ISM), and gas into the circumgalactic medium (CGM) which can result in both further cooling of the \hi\ in the galactic disk or quenching of star-formation activity. A complete picture of galaxy evolution thus requires us to understand the properties of the main constituents of galaxies, i.e., \hi, H$_2$, stars, the star-formation rate (SFR), etc, as well as their dependence on the  galaxy environment, at different cosmological times. 

Over the last three decades, much progress has been made in understanding the evolution of the stars and the SFR in galaxies \citep[e.g.][]{Madau14}.  Similarly, over the last decade, the Atacama Large Millimeter/submillimeter Array and the Northern Extended Millimetre Array have yielded significant progress in our understanding of the molecular gas properties of at least the more massive galaxies at high redshifts \citep[e.g.][]{Tacconi20}. Unfortunately, the low Einstein A-coefficient of the \hii\ line, the sole tracer of the \hi\ content of galaxies, has meant that the above progress has not been mirrored in our understanding of the atomic phase. Despite deep \hii\ observations with the best radio telescopes of today, only a handful of individual galaxies have been detected in \hii\ emission out to $z \approx 0.4-0.5$ \citep[e.g.][]{Xi24,Jarvis25,Bluebird26}. It is likely to remain very difficult to measure the \hi\ content of significant numbers of galaxies at  $z \gtrsim 1$ until the advent of the Square Kilometre Array.

``Stacking'' of the \hii\ emission signals of a large number of galaxies with accurate spectroscopic redshifts that lie within the field of view of a radio interferometer offers the possibility of measuring the average \hi\ properties of galaxy populations at high redshifts \citep[e.g.][]{Zwaan00,Chengalur01,Lah07,Rhee16,Kanekar16}. Such \hii\ stacking has been used at intermediate redshifts, $z \approx 0.4$, by a number of studies in different fields to measure the average \hi\ mass and \hi\ depletion timescales of galaxies \citep[e.g.][]{Bera19,Luber25,Bianchetti25,Bianchetti25b,Arlow26}, the \hi\ scaling relations \citep[e.g.][]{Bera23b,Bianchetti25}, the \hi\ mass function of galaxies \citep[e.g.][]{Bera22,Pan25}, the gas accretion rate \citep{Bera23a}, etc. However, at higher redshifts, $z \gtrsim 1$, the only measurements of the \hi\ properties of galaxies are from the upgraded Giant Metrewave Radio Telescope (GMRT) surveys of the DEEP2 fields \citep{Chowdhury20,Chowdhury21,Chowdhury22c}. These studies have shown that \hi\ dominates the baryon content of star-forming galaxies at $z \approx 1$ \citep{Chowdhury22b}, and that the decline in the cosmic SFR density at $z < 1$ \citep[e.g.][]{Madau14} is due to insufficient gas accretion onto massive galaxies towards the end of the era of cosmic noon \citep{Chowdhury20,Chowdhury22a}. They have also yielded the first estimates at $z \approx 1$ of the \hi\ scaling relations \citep{Chowdhury22d}, the gas accretion rate \citep{Chowdhury23}, and the \hi\ mass function of galaxies \citep{Chowdhury24}.

While the DEEP2 galaxies are mostly main-sequence galaxies at $z=0.7-1.5$, they were selected to have $\rm R_{AB} \leq 24.1$. The R-band selection (corresponding to selection at rest-frame near-ultraviolet for galaxies at $z \gtrsim 1$) implies that the DEEP2 galaxy sample is biased towards blue main-sequence galaxies at $z \gtrsim 1$ \citep{Willmer06,Newman13}. Further, while the DEEP2 survey fields have outstanding spectroscopic coverage \citep{Newman13}, they do not have Hubble Space Telescope (HST) or James Webb Space Telescope (JWST) imaging, making it difficult to determine the dependence of the average \hi\ properties of high-$z$ galaxies on their morphology. It is thus of much interest to carry out a similar deep \hii\ survey, with the GMRT Band-4 receivers, of a survey field with excellent spectroscopy at these redshifts as well as HST and JWST imaging. Further, a \hii\ stacking experiment using a sample of star-forming galaxies with different selection criteria than those of the DEEP2 sample is also important, to assess whether the bias towards blue galaxies in the DEEP2 sample has any implications for their \hi\ properties. The COSMOS field \citep{Scoville07} is the only deep extragalactic field that meets the above requirements.

 The COSMOS field has deep HST Advanced Camera for Surveys (ACS) imaging, as well as outstanding, multi-wavelength imaging from X-ray to radio wavelengths, of a contiguous 1.7~sq.~deg. region of the sky. The central region of COSMOS has also been observed with the JWST in the COSMOS-Web project \citep{Casey23}, with deep four-band NIRCAM imaging of a $0.54$~sq. deg. region and MIRI imaging of a $0.19$~sq. deg. region. Finally, the COSMOS field has excellent spectroscopic coverage out to $z=1.5$ \citep[with selection criteria different from that of the DEEP2 galaxies; e.g.][]{Lilly09,Hasinger18,Ratajczak26,Khostovan26}, providing large galaxy samples in the redshift range of our interest ($z\approx 0.7-1.5$) for \hii\ stacking, while the near-panchromatic photometry yields reliable estimates of the galaxy stellar and star-formation properties \citep[e.g.][]{Weaver22}. 

We have hence used the upgraded GMRT to carry out the Cold-H{\sc i} AT $z\approx 1$ (CAT$z$1) \hii\ survey of COSMOS (the GMRT CAT$z$1-COSMOS survey), aiming to measure the average \hi\ properties of galaxies during cosmic noon. In this {\it Letter}, we report the first detection of the stacked \hii\ emission signal from star-forming galaxies at $z \approx 1$ in the COSMOS field.\footnote{We assume a $\Lambda$ Cold Dark Matter ``737'' cosmology, with $\Omega_m = 0.3$, $\Omega_\Lambda=0.7$ and H$_0 = 70$~km~s$^{-1}$~Mpc$^{-1}$, throughout this paper.}

\begin{figure}
    \centering
    \includegraphics[width=0.6\linewidth]{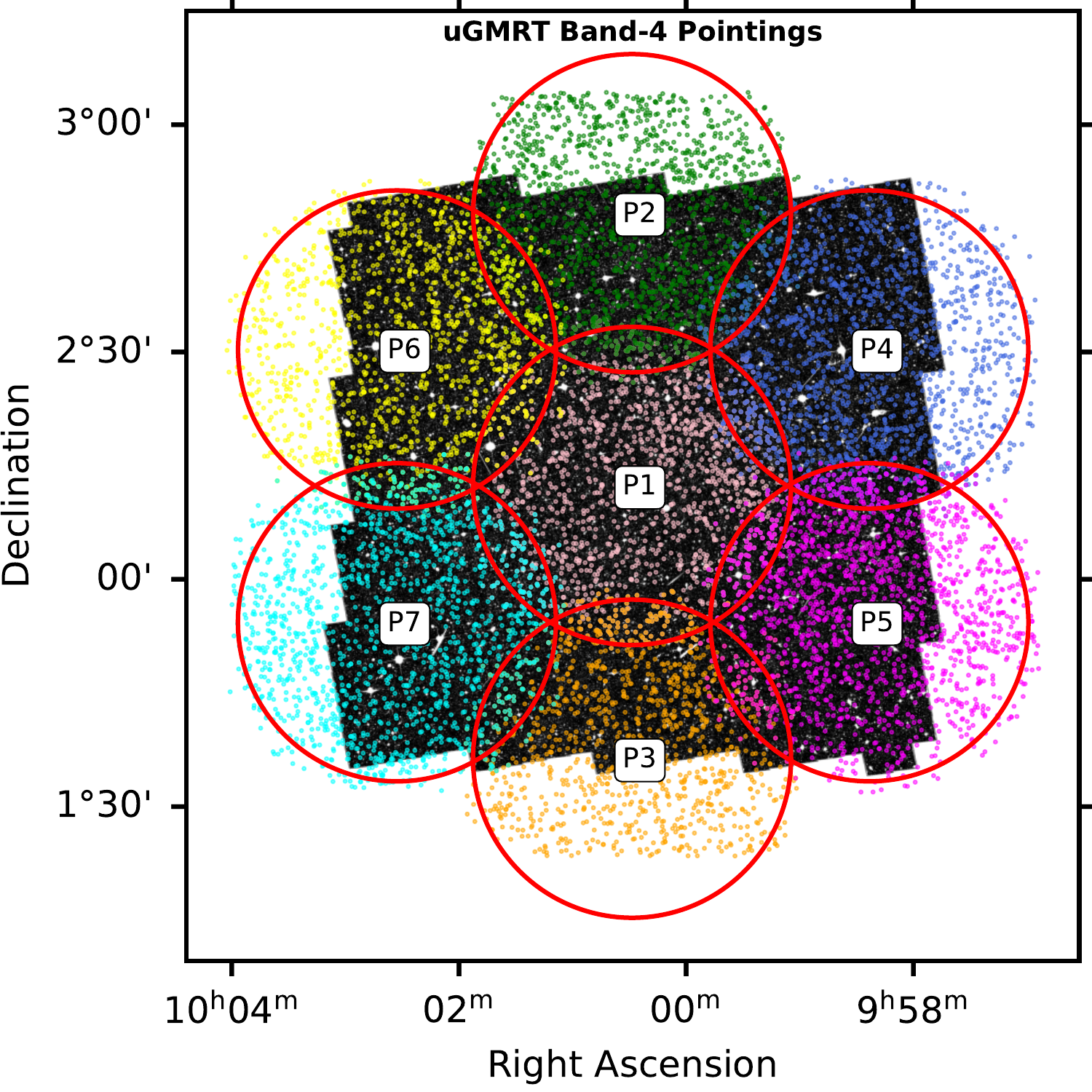}
    \caption{The figure shows the seven GMRT Band-4 pointings (labelled P1--P7) overlaid on the HST-ACS F814W image of the COSMOS field \citep[in greyscale;] []{Scoville07,Koekemoer07}. The FWHM of the GMRT Band-4 primary beam at 650~MHz, $0.7^\circ$, is indicated by the red circle around each pointing centre. In each pointing, the colored points indicate the locations of star-forming galaxies (excluding AGNs) with $\Ms \geq 10^9 \, \Msun$ at $z= 0.875-1.50$ and with accurate redshifts $Q_f=4$ \citep{Khostovan26} from the DESI, zCOSMOS-Bright, C3R2, and DEIMOS-10K surveys.}
    \label{fig:pointGMRT}
\end{figure}

\section{Observations and Data Analysis}
\label{sec:obs}

We observed the COSMOS field with the GMRT Band-4 receivers \citep{Gupta17}, covering the frequency range $550-850$~MHz. All observations used the GMRT Wideband Backend as the correlator, with a bandwidth of 400~MHz, subdivided into 16,384 channels. This yielded a native frequency resolution of $24.41$~kHz, corresponding to a velocity resolution of $9-14$~\kmps\ across the observing band. 

The GMRT CAT$z$1-COSMOS survey used two separate observing strategies. In the first approach (proposals 40\_091, 41\_058; PI: N. Kanekar), we covered the entire COSMOS field using seven different GMRT pointings, distributed in a hexagonal pattern. The separation between the pointing centres was chosen to be $0.86 \times \; {\rm FWHM}$, where FWHM~$= 0.7^\circ$ is the FWHM of the GMRT primary beam at 650~MHz \citep[e.g.][]{Condon98}. Figure~\ref{fig:pointGMRT} shows the 7 GMRT pointings overlaid on the HST ACS F814W image which covers the central 1.7~deg$^2$ region of the COSMOS field. The total sky area covered by the GMRT pointings, down to the 50\% point of the primary beam, is $2.4$~sq.~degrees. These observations extended from May~2021 to March 2022, with a total time of 133~hours. The on-source time for each of the seven pointings was $10-15$~hours.

In the second approach (proposal 45\_127; PI: N. Kanekar), we focussed on the central pointing (P1) of Figure~\ref{fig:pointGMRT}, which covers the COSMOS-Web region. Here, we use the observations carried out between December~2023 and January~2024, with a total time of $83$~hours, and an on-source time of $57$~hours.

In each observing run (typically of $5-9$-hr duration), the standard calibrators 3C147 and 3C286 were observed at the start and the end. The compact sources 0943-083 and 1041+061 were used as secondary phase calibrators; each 45m scan on COSMOS was sandwiched between 6m scans on one of these sources. The data of 3C147, 3C286, and 0943-083 were also used to calibrate the antenna bandpass shapes.

The GMRT data were analysed in the Common Astronomy Software Applications package \citep[{\sc casa} versions 5.6 and 6.6; ][]{casa07,casa22}, using standard procedures \citep[e.g.][]{Chowdhury20,Chowdhury22c}. For each observing epoch, after initially excising non-working antennas, we used the routines {\sc gaincalR} and {\sc bandpassR} \citep{calR} to determine the complex antenna-based gains and antenna-based bandpass shapes from the data on the flux density and phase calibrators. The process was carried out iteratively, with intermediate excision of outliers, until the gains and bandpasses were found to be ``clean''. The final antenna-based gains and bandpasses were next applied to the COSMOS visibilities. For each pointing except P1, the calibrated data  were then combined into a single measurement set. For P1, the data from proposals 40\_091 and 41\_058, and from proposal 45\_127, were combined into separate measurement sets.

Next, we carried out shallow flagging of the COSMOS visibilities, using the software package {\sc aoflagger} \citep{Offringa12}, to excise any strong RFI. The visibilities were then channel-averaged to a frequency resolution of $0.488$~MHz, to reduce the data volume for continuum imaging and self-calibration. The short GMRT baselines (lengths~$< 1$~km) were entirely excised at this stage, as these are typically more affected by RFI. We initially made a continuum image of size $1.5^{\circ} \times 1.5^{\circ}$, using the {\sc casa} routine {\sc tclean}, with w-projection \citep{Cornwell08}, multi-term multi-frequency synthesis with 3 terms \citep{Rau11}, and Briggs weighting \citep{Briggs95} with a robust parameter of $-0.5$. Next, we carried out a standard iterative self-calibration and imaging procedure, with typically $4-5$ rounds of phase-only self-calibration and $1-2$ rounds of amplitude-and-phase self-calibration, until the image showed no improvement on further self-calibration. We then subtracted out the continuum model from the calibrated visibilities using the routine {\sc uvsub}, and used {\sc aoflagger} again to excise low-level RFI from the residual visibilities. This was followed by a final amplitude-and-phase self-calibration and imaging cycle, to obtain the final continuum image. For the $10-15$~hrs of on-source data on pointings P1--P7 from proposals 40\_091 and 41\_058, the final imaging was done with robust=0, w-projection, and multi-term multi-frequency synthesis. For the deeper (57-hr) run on P1 in proposal 45\_127, we also used multi-scale imaging to make the final continuum image, with 3 scales corresponding to a point source, the FWHM of the synthesized beam, and three times the FWHM of the synthesized beam. The typical synthesized beam of each continuum image has FWHM~$= 4\farcs5 \times 3\farcs5$, and the rms noise on these images in each pointing is $9-12 \, \mu$Jy~beam$^{-1}$ for the 10--15-hr runs, and $6 \, \mu$Jy~beam$^{-1}$ for the 57-hr run (see Table~\ref{tab:point}).

We applied the final antenna-based gains obtained from the self-calibration procedure to the COSMOS visibilities at the original spectral resolution, and then used {\sc uvsub} to subtract out the final continuum image. We then did a final excision of low-level RFI by running {\sc aoflagger} on the residual visibilities. The higher frequencies in the band were found to be severely affected by RFI and were hence entirely flagged out. The usable frequency range is hence $565-765$~MHz, corresponding to the redshift range $z = 0.875 - 1.50$ for the \hii\ line. The fraction of visibilities that were excised in relatively ``clean'' parts of the observing band was typically $\approx 55$\%, mostly due to non-working antennas, removal of the central-square baselines, ionospheric scintillation, and RFI.

Following this, we used {\sc tclean} to make a spectral cube for each of the pointings, covering an angular extent of $0.9^\circ \times 0.9^\circ$ and the frequency range $565-765$~MHz, at a frequency resolution of 24.41~kHz. We note that the data from the different pointings were not mosaiced together to form a single joint cube, as the GMRT cubes are known to be affected by non-Gaussian systematic effects below the half-power point \citep{Chowdhury22c}. Such mosaicing would result in the joint cube containing systematic effects arising below the half-power point of each pointing. In addition, there is no advantage to mosaicing in such stacking analyses, as the galaxies are not being detected individually and so there is no possibility of a joint deconvolution. The cubes for each pointing were made using the w-projection algorithm and Briggs weighting with robust=+1. The choice of robust parameter was based on optimizing between the sensitivity and the resolution of the cubes; lower robust values also downweight the shorter baselines, reducing the effects of RFI \citep{Chowdhury22c}.  The spectral cubes have synthesized beam FWHMs $\lesssim 6\farcs5$ at 650~MHz, and rms noise $=195-527\;\mu$Jy~beam$^{-1}$ per 24.41~kHz channel. Table~\ref{tab:point} provides a summary of the observational details and the final data products.

\begin{table}
\centering
\caption{Summary of observations and results. The columns are (1)~the pointing identifier, P1--P7, (2)~the J2000 coordinates, (3)~the on-source time, in hrs, (4)~the synthesized beam of the continuum images, (5)~the rms noise on the continuum image, in  $\mu$Jy~beam$^{-1}$, (6)~the synthesized beam FWHM of the spectral cubes, at 650~MHz, (7)~the rms noise per 24.41~kHz channel on the spectral cubes, in $\mu$Jy~beam$^{-1}$ at 650~MHz, (8)~the number of galaxies from the DESI, DEIMOS-10K, zCOSMOS-Bright, and C3R2 surveys with reliable redshifts ($Q_f=4$) in the range $z = 0.875-1.50$, and within the FWHM of the GMRT Band-4 primary beam at the redshifted \hii\ line frequency, (9)~the number of independent subcubes in each field whose redshifted \hii\ emission spectra were included in the final \hii\ stack, and (10)~the median rms noise, in $\mu$Jy~Beam$^{-1}$ per 30~\kmps\ velocity channel, on the \hii\ spectra of the final sample of galaxies in each pointing, corrected for the GMRT primary beam.}
\label{tab:point}
    \begin{tabular}{c c c c c c c c c c}
    \hline
    \hline
Pointing & RA, Decn. (J2000) & Time & Beam$_{\rm cont}$ & rms$_{\rm cont}$   & Beam$_{\rm cube}$ & rms$_{\rm cube}$ & N$_{\rm gal}$ & N$_{\rm stack}$ & rms$_{\scriptsize{\textrm{H\textsc{i}}}}$\\
         & &     hrs          &   $'' \times ''$ & $\mu$Jy~beam$^{-1}$ & $'' \times ''$ & $\mu$Jy~beam$^{-1}$ & & & $\mu$Jy~beam$^{-1}$  \\
\hline
P1       & $\rm 10^h00^m28.6^s$, $+02^{\circ}12'21''$ & 10.0 & $4.7 \times 3.5$ & 11 & $6.2 \times 4.4$ & 291 & 2228 & 1597 & 525 \\
                                                    & & 57.0 & $4.2 \times 3.5$ &  6 & $5.5 \times 4.9$ & 195 & 2228 & 1566 & 269 \\
P2       & $\rm 10^h00^m28.6^s$, $+02^{\circ}48'21''$ & 15.6 & $4.3 \times 3.4$ &  9 & $5.8 \times 4.4$ & 248 & 2170 & 1456 & 360 \\
P3       & $\rm 10^h00^m28.6^s$, $+01^{\circ}36'21''$ & 14.6 & $4.1 \times 3.5$ & 12 & $5.5 \times 4.3$ & 327 & 1734 & 1129 & 441 \\
P4       & $\rm 09^h58^m23.2^s$, $+02^{\circ}30'21''$ & 15.5 & $4.2 \times 3.9$ & 10 & $5.9 \times 5.0$ & 267 & 2034 & 1329 & 402 \\
P5       & $\rm 09^h58^m23.2^s$, $+01^{\circ}54'21''$ & 15.2 & $4.6 \times 3.5$ & 10 & $6.3 \times 4.6$ & 290 & 2131 & 1335 & 392 \\
P6       & $\rm 10^h02^m32.8^s$, $+02^{\circ}30'21''$ & 14.0 & $4.4 \times 4.1$ & 12 & $6.1 \times 4.6$ & 527 & 1730 & 973  & 581 \\
P7       & $\rm 10^h02^m32.8^s$, $+01^{\circ}54'21''$ & 14.2 & $4.6 \times 3.7$ & 11 & $6.5 \times 4.6$ & 333 & 2022 & 1337 & 459 \\
\hline
\hline
\end{tabular}
\end{table}

\section{The Galaxy Sample}
\label{sec:sample}

An \hii\ stacking experiment critically requires a large number of galaxies with accurately known positions and redshifts within the primary beam of the interferometer. The COSMOS field has been a target of several spectroscopic surveys carried out using a variety of telescopes \citep[e.g.,][]{Lilly09, vanderwel21, Ratajczak26}. Very recently, \citet{Khostovan26} have provided a catalog of spectroscopic redshifts in a 10~sq.~deg. region centered on COSMOS, via the first data release of the COSMOS Spectroscopic Redshift Compilation, with spectroscopic redshifts for more than a quarter of a million galaxies out to $z \approx 8$. However, the different programs from which the catalog was compiled were carried out with a variety of telescopes, with different observational parameters, yielding very different redshift accuracies. Given that redshift accuracy is critical for our \hii\ stacking experiment, we chose to restrict our sample to emission-line galaxies from four surveys with a high expected redshift accuracy, the DESI \citep{Ratajczak26}, DEIMOS-10K \citep{Hasinger18}, zCOSMOS-Bright \citep{Lilly09}, and C3R2 \citep{Masters19} surveys, restricting to the highest-quality redshifts, $Q_f=4$, for each survey. We note that the DESI sample included in this paper is from the latest DESI data release \citep[DR2; ][]{Khostovan26,Ratajczak26}.

There are 12,323 galaxies from the above 4 surveys in our redshift range, $z=0.875-1.50$, with $Q_f=4$, and lying within the FWHM of the GMRT Band-4 primary beam at the redshifted \hii\ line frequency, of which 10,490 redshifts are from the DESI survey \citep{Ratajczak26}, 1000 from the DEIMOS-10K survey \citep{Hasinger18}, 807 from the C3R2 survey \citep{Masters19}, and 26 from the zCOSMOS-Bright survey \citep{Lilly09}. The number of galaxies with spectroscopic redshifts in each of the 7 GMRT pointings is listed in Table~\ref{tab:point}.

We emphasize that the selection of galaxies from the above four surveys implies an extremely heterogenous target selection. In our redshift range of interest, $0.875 \lesssim z \lesssim 1.50$, DESI's main galaxy samples are the Luminous Red Galaxies and the Emission Line Galaxies \citep[e.g.][]{Ratajczak26}, zCOSMOS-Bright targetted galaxies with $i \leq 22.5$ \citep{Lilly09}, DEIMOS-10K used a mixed selection, including ultraluminous infra-red galaxies, narrow- or intermediate-band excess objects, $z$-band magnitude-limited ``filler'' targets, etc \citep{Hasinger18}, while C3R2 used a complex selection to sample the full color space of galaxies down to $i<24.5$ \citep{Masters17,Masters19}. The heterogeneity of the galaxy sample has the advantage that the results of the \hii\ stacking are unlikely to have a bias stemming from the selection criteria. However, we note that the final sample is dominated by redshifts from the DESI survey.

Since the DESI survey contains the largest number of galaxies of our sample, we estimated the overall spectroscopic redshift accuracy by (1)~comparing the redshifts measured in the DESI survey to those measured in the DEEP2 survey \citep{Newman13} for galaxies in the DEEP2 fields, and (2)~comparing the redshifts of COSMOS galaxies measured in both the DESI survey and at least one of the other three surveys. The comparison between the DESI and DEEP2 redshifts for common blue star-forming galaxies (541 galaxies with stellar masses $\geq 10^9 \, \Msun$ and excluding AGNs; see below) in the DEEP2 survey fields yielded a $1\sigma$ uncertainty on the measured galaxy redshifts of $49$~\kmps. In the second comparison, the $1\sigma$ uncertainties on the measured COSMOS galaxy redshifts (again restricting to star-forming galaxies with stellar masses $\geq 10^9 \, M_\odot$ and excluding AGNs) relative to the DESI redshifts were found to be $79$~\kmps\ (zCOSMOS-Bright; 47 common galaxies), $120$~\kmps\ (DEIMOS-10k; 189 common galaxies), and $48$~\kmps\ (C3R2; 144 common galaxies). The above $1\sigma$ uncertainties exclude galaxies with catastrophic failures, with velocity offsets $> 1000$~\kmps\ between each pair of surveys: there were no such galaxies with catastrophic failures in the DESI-DEEP2 comparison, 1 such galaxy each in the DESI-C3R2 and DESI-zCOSMOS-Bright comparisons, and 13~such galaxies in the DESI-DEIMOS-10K comparison. 

Based on the number of galaxies from the different surveys, we conclude that (1)~galaxies with catastrophic redshift failures would make up $< 1$\% of the full sample and would not significantly affect the stacking results, and (2)~the overall $1\sigma$ spectroscopic redshift accuracy of the star-forming galaxy sample is $\lesssim 120$~\kmps\ (and $\lesssim 50$~\kmps\ for the DESI and C3R2 surveys, which make up more than 90\% of the sample), sufficient for accurate alignment of the galaxy redshifts for an \hii\ stacking experiment.

Besides the redshifts, \citet{Khostovan26} also provide  the physical properties and rest-frame magnitudes for all galaxies with a crossmatch in the COSMOS2020 CLASSIC catalogue \citep{Weaver22}, obtained by using the spectroscopic redshift as an input to fits to the spectral energy distribution (SED). 
Following \citet{Khostovan26}, we will use the physical properties and rest-frame photometry derived from the {\sc CIGALE} SED fits \citep{Boquien19,Yang20}, excluding 905 galaxies in the catalogue for which this information was not  available from the {\sc CIGALE} fits. Finally, we used the NUVrJ color-color diagnostic  \citep{Ilbert13} to separate between quiescent and star-forming galaxies, where quiescent galaxies have $NUV-r > 3.1$ and $NUV-r > 3(r-J) + 1$. This color-color selection yields a sample of 10,123 unique star-forming galaxies in the seven GMRT pointings. 

Next, we excluded any galaxies containing active galactic nuclei (AGNs) from the sample, as the presence of an AGN could affect the \hii\ emission profile of the galaxy, due to both feedback effects affecting the \hi\ content and (for radio-loud AGNs) the possibility of \hii\ absorption against the AGN continuum. Note that we have already excluded galaxies classified as AGNs based on the presence of broad emission lines, consistent with an AGN, as such galaxies have $Q_f=14$ in the catalogue of \citet{Khostovan26}. We further excluded any galaxies with X-ray detections in the COSMOS2020 catalogue \citep{Weaver22}. We also excluded objects that are likely to be radio AGNs based on their detections at $\geq 4\sigma$ significance in our GMRT 680~MHz continuum images with inferred rest-frame 1.4-GHz luminosity $L_{1.4~\rm GHz} \geq 2 \times 10^{23}$~W~Hz$^{-1}$ \citep{Condon02}. 552 objects were excluded due to the likely presence of an AGN, leaving a sample of 9,504 star-forming galaxies.

We followed \citet{Chowdhury22c} in excluding 350 galaxies with low stellar masses, $\Ms < 10^9 \, \Msun$. This was done so as to allow a direct comparison between the results of our COSMOS survey with those of the GMRT CAT$z$1 survey \citep{Chowdhury22c} and the stellar mass-selected extended GALEX Arecibo SDSS Survey \citep[xGASS; ][]{Catinella18} at $z \approx 0$. After the above exclusions, our galaxy sample includes 9,154 star-forming galaxies, with $\Ms \geq 10^9 \, \Msun$.

\begin{figure}
    \centering
    \includegraphics[width=0.9\linewidth]{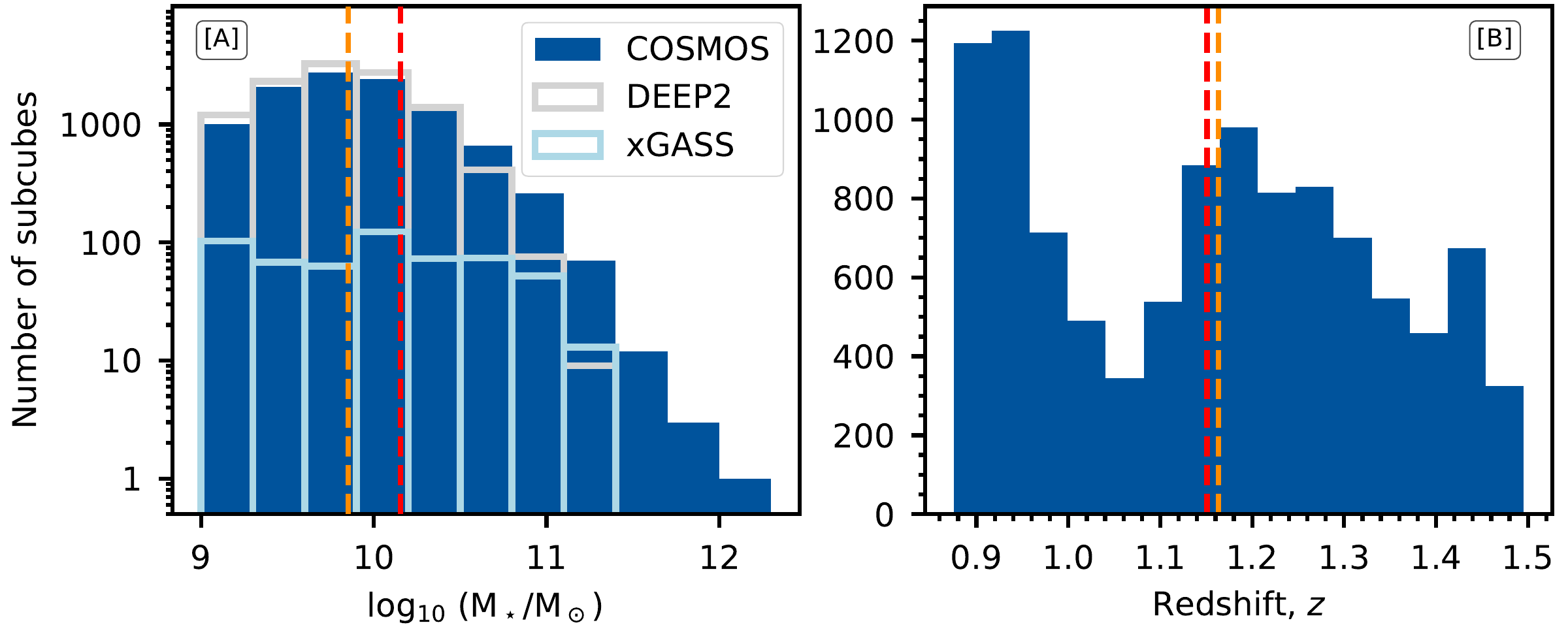}
    \caption{The distributions of [A]~the logarithm of the stellar mass, and [B]~the redshift of the 10,722 independent subcubes (from 8,122 unique galaxies), that were used in the final \hii\ stacking analysis. The dashed orange and red lines indicate the median and the mean values of each distribution, respectively. For comparison, we have also plotted in panel~[A] the stellar-mass distributions of the DEEP2 galaxies of the GMRT CAT$z1$ survey \citep[grey open histogram;][]{Chowdhury22c} and the xGASS survey \citep[light blue open histogram; ][]{Catinella18}.}
    \label{fig:distribution}
\end{figure}

\section{The \hii\ subcubes: Statistical tests }
\label{sec:subcubes}

Some of the 9,154 star-forming galaxies of the final sample lie in the regions of overlap of the GMRT pointings (see Figure~\ref{fig:pointGMRT}), and hence are covered in more than one of our spectral cubes. In addition, the central pointing P1 has two spectral cubes, the first from the short observations of proposals 40\_091 and 41\_058, and the second from the deeper observations of proposal 45\_127. 

For each of the 9,154 galaxies, we extracted a subcube of size $307\farcs2 \times 307\farcs2$ centered on the galaxy location from each spectral cube that contains the galaxy within the FWHM of the GMRT primary beam at its redshifted \hii\ line frequency. The velocity range of the subcubes was chosen to be $\pm1900$~\kmps, in the rest frame of each galaxy and centered on the galaxy redshift. We then convolved each subcube with a Gaussian beam to a uniform spatial resolution of 90~kpc at the galaxy redshift. This resolution was chosen based on the DEEP2 results of \citet{Chowdhury22c}, who found that some of the stacked \hii\ emission signal is resolved out for spatial resolutions $< 90$~kpc\footnote{We also carried out the \hii\ stacking at spatial resolutions of 60~kpc and 80~kpc, and found results consistent (within $1\sigma$ significance) with our final results at 90-kpc resolution.}. We next regridded the convolved subcubes to a uniform pixel grid of size 10~kpc and a velocity resolution of 30~\kmps. Each subcube was then normalized so that the peak of its convolved point spread function remains unity when convolved with the same Gaussian beam \citep{Chowdhury20,Chowdhury22c}. Next, we corrected the flux density of each subcube to account for the location of the galaxy  in the GMRT primary beam. Finally, we fitted a second-order polynomial to the spectrum from each spatial pixel of each subcube, and subtracted this out. The final residual subcubes have a spatial extent of $\rm 1~Mpc \times 1~Mpc$ and a velocity range of $\pm 1500$~\kmps\ in the rest frame of each galaxy, with a velocity resolution of 30~\kmps\ and a spatial resolution of  90~kpc. In total, there were 12,047 subcubes for the 9,154 unique galaxies.

Next, we carried out a set of statistical tests \citep[following][]{Chowdhury22c}, to excise galaxy subcubes affected by non-Gaussian statistics or lying at frequency ranges severely affected by RFI. Specifically, we removed any subcubes that matched any of the following criteria:

\begin{itemize}

\item Subcubes with more than 20\% of the channels flagged due to RFI. 

\item Subcubes that showed spectral features (positive or negative) with $\geq 6\sigma$ significance at velocity resolutions of 30, 120, or 300~\kmps.

\item Subcubes whose rms noise decreased by a factor $> 0.4$ on smoothing from a velocity resolution of 30~\kmps\ to 300~\kmps\ (note that the rms noise on such smoothing is expected to decrease by a factor of $\sqrt{30/300} = 0.316$ for Gaussian noise). This was done in order to excise subcubes showing correlations across multiple spectral channels.
    
\end{itemize}

After the above exclusions, there are 8,122 unique star-forming galaxies, with 10,722 independent subcubes, in our final sample. Figure~\ref{fig:distribution} shows histograms of the stellar mass and the redshift distributions of the final sample of 10,722 independent subcubes. The dashed vertical line in each panel shows the median value of each distribution, at $\Ms = 7.1 \times 10^9 \, \Msun$ and $z = 1.1635$.

\section{\hii\ and continuum stacking: The average \hi\ mass and the median SFR}

We initially converted the subcubes from flux density to  \hii\ line luminosity density $L_{\rm HI}$, using the relation $L_{\rm HI} = 4\pi\ {\rm S_{HI}D_L^2}/(1+z)$. Next, for each subcube, we measured the spectral rms noise per 30~\kmps\ channel over the entire subcube, in units of luminosity density. The inverse square of this rms noise of each subcube was used as its weight in the \hii\ stacking, to improve the signal-to-noise ratio. We then carried out a weighted average of the 10,722 subcubes, pixel by pixel, to generate the stacked \hii\ spectral cube. Finally, we fitted the spectrum at each spatial pixel of the above cube with a second-order polynomial, excluding the velocity range $\pm200$~\kmps, and subtracted this out to remove any residual spectral baseline and obtain the final  stacked \hii\ cube.

We used a Monte Carlo simulation to estimate the rms noise on the final stacked cube, carrying out 10,000 realizations of the weighted average of the 10,722 subcubes but with randomized redshifts. The rms noise on each pixel across the above 10,000 Monte Carlo realizations was used as the final rms noise on the corresponding pixel of the stacked cube \citep[e.g.][]{Chowdhury22c}. 

\begin{figure}
    \centering
    \includegraphics[width=0.9\linewidth]{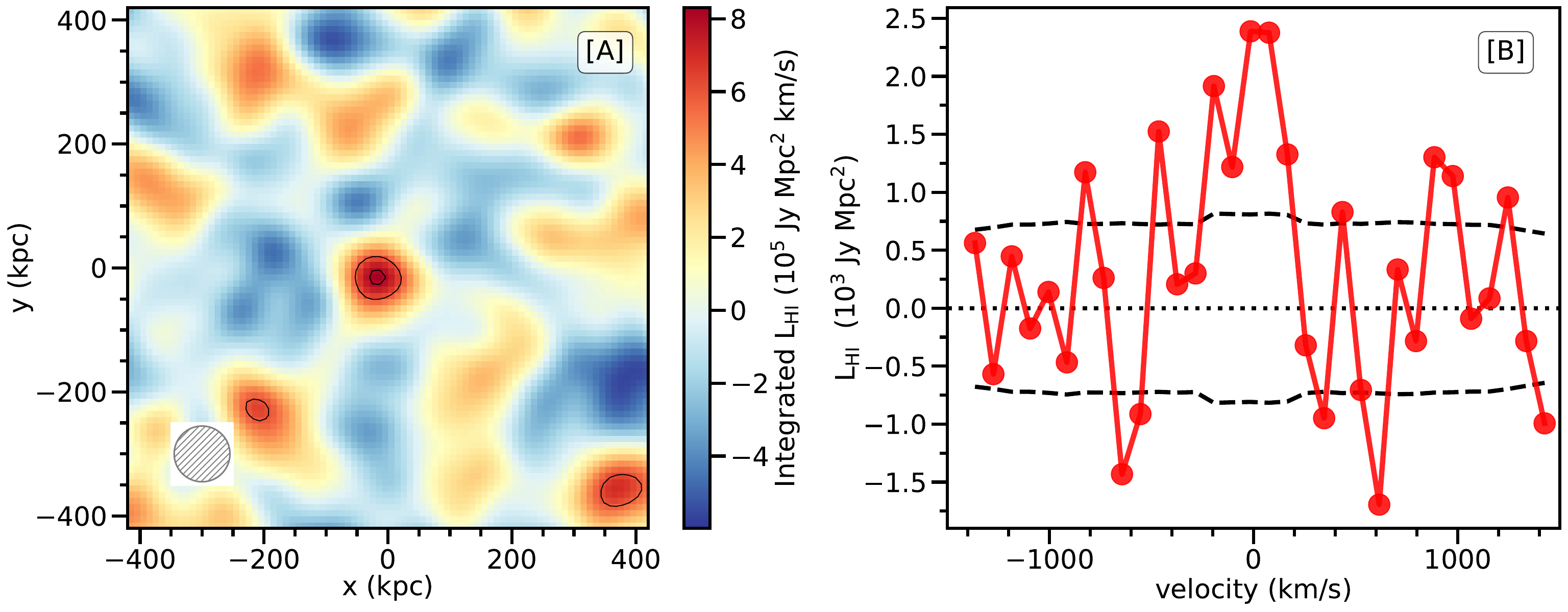}
    \caption{The stacked \hii\ emission signal from the 8,122 star-forming galaxies in COSMOS at $z= 0.875-1.50$. [A]~The velocity-integrated \hii\ moment-0 image, where the integral is over the central velocity range $[-200, +200]$~\kmps\ of the stacked \hi\ cube. The inset on the bottom left corner shows the FWHM of the beam of 90~kpc to which the individual galaxy subcubes were convolved before stacking. The contours are at $(3, 4) \times \sigma$ significance; note that there are no contours at the $-3\sigma$ level. [B]~The stacked \hii\ spectrum at a velocity resolution of 90~\kmps, extracted from the location of the peak in the \hii\ moment-0 image in [A]. The dashed black curves show the $\pm1\sigma$ rms noise per 90~\kmps\ velocity channel.}
    \label{fig:stack}
\end{figure}

Figure~\ref{fig:stack}[A] shows the \hii\ moment-0 image, obtained by integrating the final stacked \hii\ cube over the velocity range $[-200~{\rm km~s^{-1}}, +200~{\rm km~s^{-1}}]$. The spatial resolution of the image is 90~kpc, indicated by the circle at the bottom left. Figure~\ref{fig:stack}[B] shows the spectrum extracted at the location of the peak of the \hii\ moment-0 image of Figure~\ref{fig:stack}[A], after smoothing to, and resampling at, a velocity resolution of 90~\kmps. The dashed curve shows the rms noise per 90~\kmps\ channel. Both panels of Figure~\ref{fig:stack} show a clear detection of the stacked \hii\ signal, at the centre of the stacked cube. Integrating over the central $\pm 200$~\kmps, the velocity-integrated \hii\ line luminosity density is $(8.3 \pm 2.0) \times 10^5$~Jy~Mpc$^2$~km~s$^{-1}$. This corresponds to an average \hi\ mass of $\MHI = (15.5 \pm 3.7) \times 10^9~\Msun$ for the 8,122 star-forming galaxies of the final sample. The FWHM of the stacked \hii\ line is $284 \pm 33$~\kmps, from a single Gaussian fit to the profile in Figure~\ref{fig:stack}[B].

We used our 680~MHz continuum images and the correlation between the rest-frame 1.4~GHz radio luminosity and the SFR \citep[e.g.][]{Yun01} to obtain an independent estimate of the median SFR of the 8,122 galaxies of our sample. For each galaxy, we extracted a cutout image of size $84\farcs4 \times 84\farcs4$, centered on the galaxy. We convolved each cutout image with a 2-D Gaussian to obtain a uniform spatial resolution of 42~kpc \citep[far larger than the size of the radio continuum in star-forming galaxies at $z \approx 1$; e.g.][]{Jimenez-Andrade19}, and regridded each image to a uniform spatial grid made up of 5~kpc pixels. For the redshift range of our galaxies, $z = 0.875-1.50$, the 680~MHz continuum images are at rest frequencies of $1.3-1.7$~GHz. We used a spectral index of $\alpha=-0.8$ (assuming that flux density $S_\nu \propto \nu^\alpha$) and scaled each image to a rest-frame frequency of 1.4~GHz. Finally, we converted the continuum subimage of each galaxy from flux density to rest-frame 1.4~GHz luminosity, applied the weights used for the \hii\ stacking above, and carried out a weighted median stack, pixel by pixel of the subimages \citep[e.g.][]{White07}. In passing, we note that median stacking was used for the continuum images, as median stacking is less affected than mean stacking by outliers in the distribution (e.g. due to the brighter continuum sources or non-Gaussian deconvolution errors). Further, \citet{White07} have shown that, in the low S/N regime where the signals are lower than the RMS noise, the measured median converges to the mean value of the distribution, even for non-Gaussian distributions. They hence recommend median stacking for continuum emission, and we have followed this approach here. We also carried out the above median stacking procedure for locations offset by $120''$ from the 8,122 galaxies, to test for possible systematics in the stacked continuum image.

\begin{figure}
    \centering
    \includegraphics[width=0.9\linewidth]{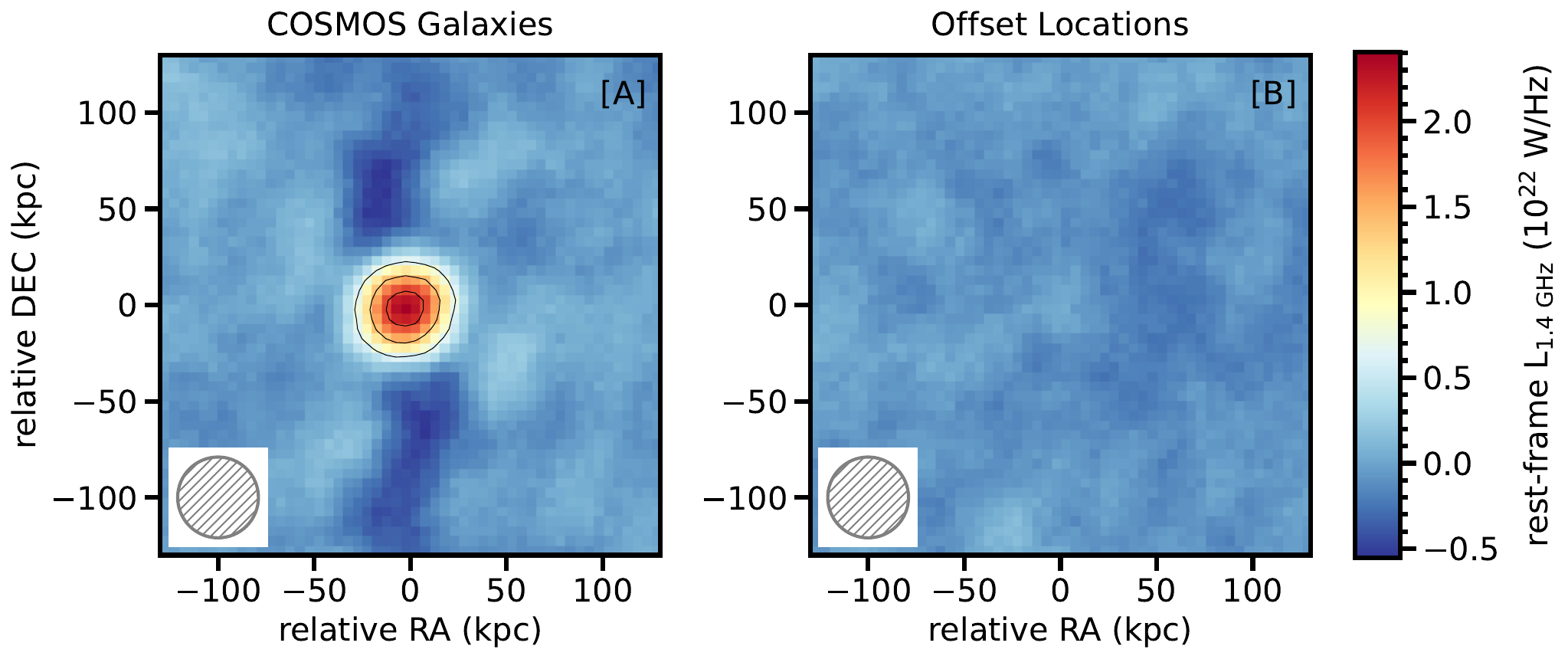}
    \caption{[A]~The weighted-median stack of the rest-frame 1.4~GHz luminosity of the final sample of 8,122 star-forming galaxies, in units of W~Hz$^{-1}$. We obtain a clear detection of the stacked rest-frame 1.4~GHz continuum emission, at $35\sigma$ significance. Note the negative features north and south of the stacked continuum source arise due to the undeconvolved point-spread function of the interferometer. [B]~The weighted-median stacked image of the 1.4~GHz luminosity at locations offset by $120''$ from each of the 8,122 galaxies; this image shows no evidence for systematic effects. The contours in [A] are plotted at $(10,20,30) \times \sigma$, where $\sigma$ is the rms noise on the stacked offset image in [B]. }
    \label{fig:continuum}
\end{figure}

The final median-stacked rest-frame 1.4~GHz continuum image of the 8,122 galaxies is shown in Figure~\ref{fig:continuum}[A]. This yields a  $35\sigma$ detection of the stacked rest-frame 1.4-GHz continuum signal, with an average rest-frame 1.4-luminosity of $L_{\rm 1.4~GHz} = (2.40 \pm 0.25) \times 10^{22}$~W~Hz$^{-1}$, where the $1\sigma$ error was obtained by adding in quadrature the measurement error and the typical 10\% uncertainty in the GMRT flux-density scale for such interferometric observations. Using the conversion ${\rm SFR} = 3.7 \times 10^{-22} \times L_{\rm 1.4~GHz}$ \citep[for a Chabrier initial mass function;][]{Yun01}, we obtain an average SFR of $(8.90 \pm 0.92)~\Msun$~yr$^{-1}$. Finally, Figure~\ref{fig:continuum}[B] shows the stacked 1.4-GHz continuum image at the offset locations, $120''$ away from the galaxies of our sample. We find that the offset stacked image exhibits noise-like behaviour, with no evidence for systematic effects.  

An important issue for a survey such as ours is whether the comoving volume of the survey is sufficiently large that the inferred average galaxy properties are representative of the galaxy population at the target redshift range, or whether the inferred properties might have significant uncertainty due to cosmic variance \citep[e.g.][]{Driver10}. We hence evaluated the effective comoving volume of the GMRT CAT$z$1-COSMOS survey, and estimated the effects of cosmic variance on our results. Each of the GMRT Band-4 pointings covers a comoving volume of $2.4 \times 10^6$~cMpc$^3$, over the redshift range $0.875 \lesssim z \lesssim 1.50$. However, the use of weights proportional to the inverse variance in luminosity density units in the \hii\ stacking implies that the effective comoving volume for each pointing is smaller than the above value, because galaxies in the outer parts of the primary beam and at higher redshifts are given lower weights in the stack. Including the effect of the above weighting, the effective comoving volume for each pointing is $4.8 \times 10^5$~cMpc$^3$. In addition, it can be seen from Table~\ref{tab:point} that the median rms noise for pointing P1 is significantly lower than that of the other pointings, due to the significantly higher observing time on this pointing. We used these median rms noise values to find that pointing P1 dominates the effective comoving volume of the full GMRT survey, contributing approximately one-third of the total effective comoving volume. Taking the sensitivities of the different pointings into account, we find that the effective comoving volume of the full GMRT survey of COSMOS is $1.4 \times 10^6$~cMpc$^3$. Note that the correction due to the overlap regions of the pointings is small, $< 10$\%. 
\citet{Driver10} provide estimates of the cosmic variance for a normal galaxy survey based on analysis of the Sloan Digital Sky Survey (SDSS) volume, at $z \lesssim 0.1$. Extending the SDSS results to higher-redshift surveys (see their Section~3.4), the upper limit to the cosmic variance on our estimate of the average galaxy properties for a contiguous volume of size $1.4 \times 10^6$~cMpc$^3$ is $\lesssim 20$\% \citep{Driver10}. We conclude that cosmic variance is not a significant source of uncertainty on our estimates of the average \hi\ properties of galaxies in the COSMOS field.

\section{Discussion}

\begin{figure}
    \centering
    \includegraphics[width=0.8\linewidth]{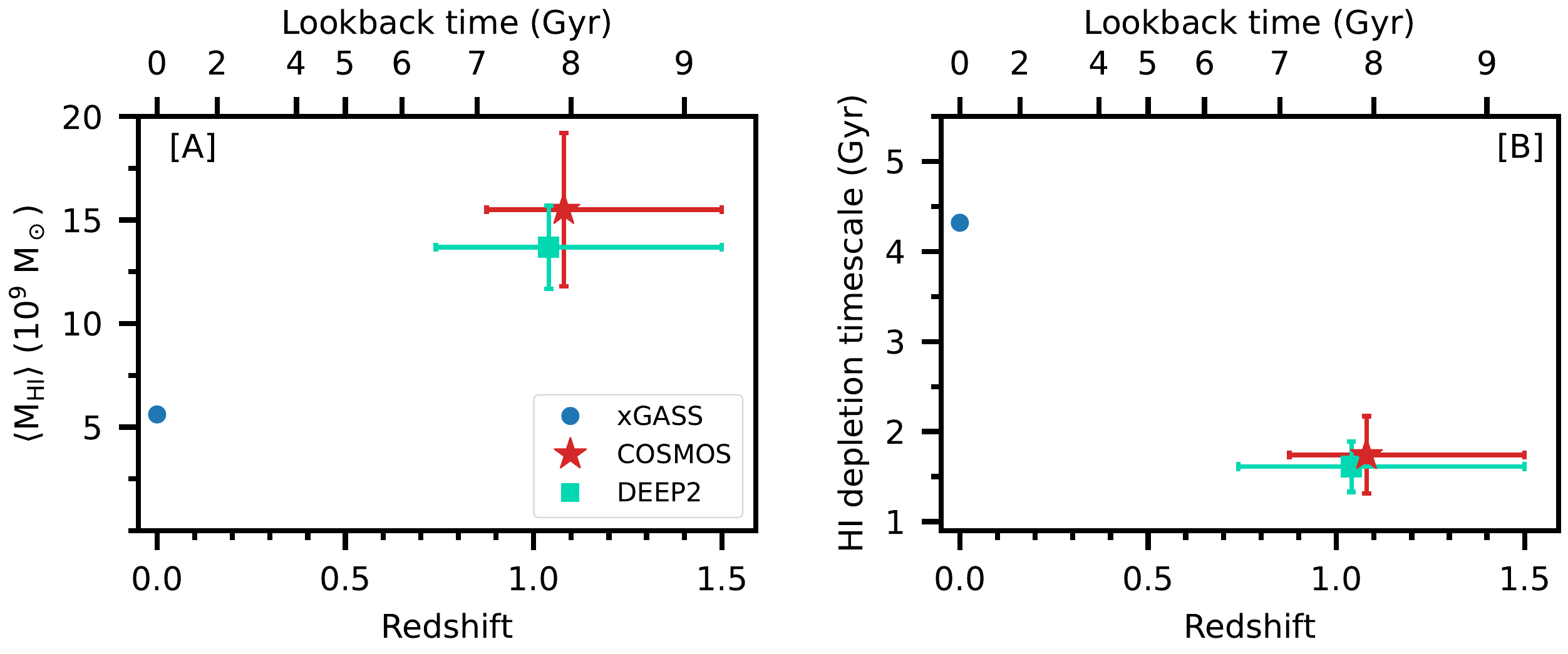}
    \caption{The plot shows a comparison between [A]~the average \hi\ mass, and [B]~the characteristic \hi\ depletion timescale, for stellar mass-matched samples of star-forming galaxies in COSMOS (red star; this work), blue star-forming galaxies in the DEEP2 survey fields \citep[green squares;][]{Chowdhury22c}, and blue star-forming galaxies from the xGASS survey \citep[blue circles;][]{Catinella18}. It is clear that both the average \hi\ mass and the characteristic \hi\ depletion timescale are in excellent agreement for the COSMOS and DEEP2 galaxies at $z\approx1$, and that both are very different from the corresponding values for the xGASS galaxies at $z\approx0$.}
    \label{fig:mhi_tau}
\end{figure}

Until the present work, all of our information on the \hi\ properties of star-forming galaxies at $z\gtrsim1$ has come from the GMRT \hii\ surveys of the DEEP2 fields \citep{Chowdhury20,Chowdhury21,Chowdhury22c}. Figure~\ref{fig:stack} presents only the second detection of the stacked \hii\ emission signal from galaxies at $z \approx 1$. Further, while the DEEP2 galaxies have a ``clean'' selection criterion, $\rm R_{AB} \leq 24.1$ \citep{Newman13}, the rest-frame near-ultraviolet selection for galaxies at $z\gtrsim 1$ makes the DEEP2 sample at $z \gtrsim 1$ biased towards blue star-forming galaxies \citep{Willmer06}. Conversely, while the COSMOS galaxies stacked here were selected from four different surveys with different selection criteria, and thus constitute a heterogenous sample, the outstanding multi-wavelength coverage in COSMOS allowed us to select star-forming galaxies based on the NUVrJ color-color criterion \citep{Ilbert13}. In this section, we will compare our COSMOS results for star-forming galaxies at $z \approx 1$ to the results of \citet{Chowdhury22c} for blue star-forming galaxies at $z \approx 1$ in the DEEP2 fields, and \citet{Catinella18} for the stellar mass-selected sample of xGASS galaxies at $z \approx 0$.

First, we tested for possible biases in our final COSMOS galaxy sample (e.g. towards higher SFRs) by measuring the median SFR of subsamples of galaxies in multiple stellar-mass bins and in two redshift bins, $z=0.875-1.15$ and $z=0.875-1.50$. We find that, for each redshift and stellar-mass bin, the median SFRs of our galaxies are in excellent agreement with the star-forming galaxy main-sequence relation of \citet{Popesso23}. This indicates that the final COSMOS galaxy sample is drawn from the main sequence at $z \approx 0.875-1.50$.

Using the weights applied in the \hii\ stacking, our sample of star-forming galaxies in COSMOS has a mean stellar mass of  $\langle \Ms \rangle =1.35 \times 10^{10}~\Msun$, at a mean redshift of $\langle z \rangle = 1.081$. Our GMRT \hii\ and 1.4-GHz radio continuum stacking yielded, respectively, an average \hi\ mass of $\MHI= (15.5 \pm 3.7) \times 10^9~\Msun$ and an average SFR of $(8.90\pm0.92)~\Msun~\rm yr^{-1}$. We thus obtain a characteristic \hi-to-stellar mass ratio of $\MHI/\langle \Ms \rangle = (1.15 \pm 0.27)$ and a characteristic \hi\ depletion timescale of $\MHI/{\rm SFR} = (1.74\pm0.45)$~Gyr for star-forming galaxies at $\langle z \rangle = 1.081$.

Next, we compared our results with those for the blue star-forming galaxies in the DEEP2 fields \citep{Chowdhury22c} and the blue xGASS galaxies at $z \approx 0$ \citep{Catinella18}. First, we note that both these samples are made up of mostly main-sequence galaxies at their respective redshifts \citep{Chowdhury22c}: we are thus comparing the \hi\ properties of main-sequence galaxies in COSMOS at $z \approx 1$ with those of main-sequence galaxies in the DEEP2 fields at $z \approx 1$, and in the xGASS sample at $z \approx 0$. We measured the average \hi\ mass in the DEEP2 and xGASS galaxies  using weights such that the stellar-mass distribution in both samples matches that of our COSMOS galaxies, after accounting for the luminosity-density weights used in our stacks.\footnote{The intrinsic stellar-mass distributions of the DEEP2 galaxies at $z \approx 0.74-1.45$ of the GMRT~CAT$z$1 survey \citep{Chowdhury22c} and the $z \approx 0$ galaxies of the xGASS survey \citep{Catinella18} are shown in open histograms in Figure~\ref{fig:distribution}[A]. We note that there are 15 galaxies in the COSMOS sample that have stellar masses higher than the highest stellar mass of the DEEP2 and xGASS samples. Since the number of such galaxies is small, there is no significant change in the average \hi\ mass when retaining them; we hence did not drop these galaxies from the COSMOS sample in the comparison with the DEEP2 and xGASS samples.} This yielded average \hi\ masses of $\MHI_{\rm DEEP2} = (13.3 \pm 2.1) \times 10^{9} \, \Msun$ at $\langle z \rangle = 1.04$ and $\MHI_{\rm xGASS} = (5.61 \pm 0.26) \times 10^9 \, \Msun$ at $z\approx 0$ for the DEEP2 and xGASS galaxies, respectively. The corresponding average SFRs are ${\rm SFR}_{\rm DEEP2} = (8.63 \pm 0.86)\, \Msun$~yr$^{-1}$ and ${\rm SFR}_{\rm xGASS} =  1.30 \, \Msun$~yr$^{-1}$, respectively. This implies characteristic \hi-to-stellar mass ratios of $(1.02 \pm 0.16)$ (DEEP2) and $0.42$ (xGASS), and characteristic \hi\ depletion timescales of $(1.54 \pm 0.29)$~Gyr (DEEP2) and $4.3$~Gyr (xGASS). In passing, we note that the average SFRs of the stellar mass-matched COSMOS and DEEP2 galaxy samples are in excellent agreement, despite the very different selection criteria: both galaxy samples are consistent with the main sequence at $z \approx 1$ \citep[e.g.][]{Chowdhury22c}.

Figure~\ref{fig:mhi_tau}[A] plots the average \hi\ mass of the COSMOS, DEEP2, and xGASS galaxies against redshift. We note that all three samples have matched stellar-mass distributions, and very similar average stellar masses. It is clear from the figure that the average \hi\ mass of the star-forming galaxies in COSMOS is consistent, within the statistical uncertainties, with that of the blue star-forming galaxies in the DEEP2 fields, and that both are $\approx 2.5$ times higher than the average \hi\ mass of the xGASS galaxies at $z \approx 0$. We thus find independent evidence that star-forming galaxies at $z \approx 1$ are significantly more gas-rich than at $z \approx 0$, for the same stellar-mass distribution.

Figure~\ref{fig:mhi_tau}[B] plots the characteristic \hi\ depletion timescale of the COSMOS, DEEP2, and xGASS galaxies against redshift. Again, it is clear that the characteristic \hi\ depletion timescales of the COSMOS and DEEP2 galaxies are in excellent agreement with each other, and that both are $\approx 2.5$ times smaller than the \hi\ depletion timescale of the xGASS galaxies at $z \approx 0$. As emphasized by \citet{Chowdhury20,Chowdhury22a}, the short \hi\ depletion timescale for star-forming galaxies at $z \approx 1$ provides an explanation for the decline in the SFR density of the Universe at $z < 1$, if the \hi\ accretion from the circumgalactic medium and the intergalactic medium is insufficient to replenish the \hi\ reservoir consumed in the process of star-formation.
\begin{table}

\caption{Results of the \hii\ and continuum stacking analyses. The average redshifts and stellar masses, and median rest-frame 1.4~GHz luminosities, and SFRs, are all weighted quantities, with the same weights as those used for the \hii\ stacking. }
\centering
\label{tab:summary}
    \begin{tabular}{| c  | c |}
    \hline

\hline
\hline

Number of unique Galaxies       &  8,122\\
\hline
Number of \hii\ subcubes       & 10,722\\
\hline
Redshift range           &   $0.875-1.496$       \\
\hline
Average redshift, $\langle z \rangle$         & 1.081 \\
\hline
Stellar Mass range, $\Ms (\Msun)$             & $1 \times 10^{9} - 1.9 \times 10^{12}$\\
\hline
Average Stellar mass, $\langle \Ms \rangle$ ($\Msun$) & $1.35 \times 10^{10}$\\
\hline
Median rest-frame $L_{1.4 \rm GHz}, \rm (W~Hz^{-1})$ & $(2.40 \pm 0.25) \times 10^{22}$\\
\hline
Average SFR ($\Msun/ \rm yr^{-1}$)       & $(8.90 \pm 0.92)$\\
\hline
Average \hi\ mass, $\MHI$ ($\Msun$)       & $(15.5 \pm 3.7) \times 10^9 $\\
\hline
Characteristic \hi-to-stellar mass ratio, $\MHI/\langle \Ms \rangle$ & $(1.15 \pm 0.27)$ \\
\hline
Characteristic \hi\ depletion time, $\MHI/SFR$ (Gyr)       & $(1.74 \pm 0.45) $\\

\hline
    \end{tabular}
\end{table}
\section{Summary} 
\label{sec:summary}

We report results from the GMRT CAT$z$1-COSMOS survey, a deep Band-4 observation of the COSMOS field, aimed at characterizing the \hi\ properties of star-forming galaxies at $z \approx 1$. The GMRT observations used a 7-pointing strategy to cover a $2.4$~sq.~degree region in COSMOS, combined with a deeper observation of the central COSMOS-Web region; the total observing time was $216$~hrs. Our final galaxy sample included 8,122 star-forming galaxies at $z = 0.875-1.50$, selected using the NUVrJ color-color criterion, and with accurate spectroscopic redshifts (redshift quality $Q_f=4$ and $1\sigma$ redshift uncertainty $< 120$~\kmps) from the DESI, DEIMOS-10K, C3R2, or zCOSMOS-Bright surveys. The heterogeneous selection implies that the galaxy sample is unlikely to contain an explicit bias stemming from the selection criteria, although the sample is dominated by galaxies from the DESI survey.

We stacked the \hii\ emission signals of the 8,122 star-forming galaxies at a spatial resolution of 90~kpc, measuring an average \hii\ line luminosity of $(8.3 \pm 2.0) \times 10^5$~Jy~Mpc$^2$~km~s$^{-1}$. This is only the second detection of the stacked \hii\ emission signal from galaxies at $z \approx 1$, after the detections of \citet{Chowdhury20,Chowdhury21,Chowdhury22c} in the DEEP2 survey fields. We obtain an average \hi\ mass of $\MHI = (15.5 \pm 3.7) \times 10^9 \, \Msun$ and an average SFR of $(8.90\pm0.92) \, \Msun$~yr$^{-1}$ for the 8,122 galaxies of the sample. The effective comoving volume of the GMRT survey is $1.4 \times 10^6$~cMpc$^3$, taking into account the inverse-square weighting by the rms noise in luminosity density units and the fact that significantly more time was spent on the central pointing. This implies an upper limit of $20$\% to the uncertainty on the above estimates of the average \hi\ mass and the average SFR due to cosmic variance.

We obtain a characteristic \hi-to-stellar mass ratio of $(1.15 \pm 0.27)$ and a characteristic \hi\ depletion timescale of $(1.74\pm0.45)$~Gyr, for star-forming galaxies at $\langle z \rangle = 1.081$. Our estimates of the characteristic \hi-to-stellar mass ratio and the \hi\ depletion timescale are in good agreement with the corresponding values for a stellar mass-matched sample of blue star-forming galaxies at $z \approx 1$ from the DEEP2 survey. We find that, like the blue star-forming galaxies in the DEEP2 fields, the star-forming galaxies in COSMOS at $z \approx 1$ are significantly more gas-rich than a stellar mass-matched sample of blue star-forming galaxies at $z \approx 0$ from the xGASS survey. Further, we find that star-forming galaxies at $z \approx 1$ also have far shorter \hi\ depletion timescales than star-forming galaxies at $z \approx 0$. The short \hi\ depletion timescale in massive star-forming galaxies at $z\approx1$ provides an explanation for the decline in the SFR density of the  Universe after the era of cosmic noon. Finally, we emphasize that our results and conclusions are now supported by two independent, and statistically consistent, measurements of the average \hi\ properties of star-forming galaxies at $z \approx 1$, in two different survey fields with entirely different galaxy selection criteria, each covering large comoving volumes such that cosmic variance is unlikely to play a significant role.

\begin{acknowledgements}
We thank an anonymous referee for a very detailed, careful, and insightful report that improved this Letter. We thank the staff of the GMRT who have made these observations possible. The GMRT is run by the National Centre for Radio Astrophysics of the Tata Institute of Fundamental Research. We acknowledge support from the Department of Atomic Energy, under project 12-R\&D-TFR-5.02-0700, and project RTI4017 ``Next Generation Instrumentation for Radio Astronomy''. NK also acknowledges support from the Science and Engineering Research Board of the Department of Science and Technology via a J. C. Bose Fellowship (JCB/2023/000030). 
\end{acknowledgements}

\facilities{GMRT}

\software{{\sc casa} \citep{casa22}; {\sc calR} \citep{Chowdhury21}.}
\bibliographystyle{aasjournalv7}
\bibliography{bibliography}

\end{document}